\documentclass[unsortedaddress,pre,amsmath,amssymb]{revtex4}
\usepackage[dvips]{graphicx}
\usepackage{dcolumn}
\usepackage{bm}
\usepackage{graphicx}  
\begin{document}
\title{Traffic phenomena in biology:
from molecular motors to organisms{\footnote{Based on a plenary lecture delivered by DC at TGF05, Berlin.}} }
\author{Debashish Chowdhury}
\affiliation{Department of Physics, Indian Institute of Technology,
Kanpur 208016, India.} 
\author{Andreas Schadschneider}
\affiliation{Institut  f\"ur Theoretische  Physik, Universit\"at
zu K\"oln D-50937 K\"oln, Germany}
\author{Katsuhiro Nishinari}
\affiliation{Department of Aeronautics and Astronautics,
Faculty of Engineering, University of Tokyo,
Hongo, Bunkyo-ku, Tokyo 113-8656, Japan.}
%

\begin{abstract}
Traffic-like collective movements are observed at almost all levels
of biological systems. Molecular motor proteins like, for example,
kinesin and dynein, which are the vehicles of almost all
intra-cellular transport in eukayotic cells, sometimes encounter
traffic jam that manifests as a disease of the organism. Similarly,
traffic jam of collagenase MMP-1, which moves on the collagen fibrils
of the extracellular matrix of vertebrates, has also been observed
in recent experiments. Traffic-like movements of social insects like
ants and termites on trails are, perhaps, more familiar in our everyday
life. Experimental, theoretical and computational investigations in the
last few years have led to a deeper understanding of the generic or
common physical principles involved in these phenomena. In particular,
some of the methods of non-equilibrium statistical mechanics, pioneered
almost a hundred years ago by Einstein, Langevin and others, turned out
to be powerful theoretical tools for quantitative analysis of models of
these traffic-like collective phenomena as these systems are
intrinsically far from equilibrium. In this review we critically examine
the current status of our understanding, expose the limitations of the
existing methods, mention open challenging questions and speculate on
the possible future directions of research in this interdisciplinary
area where physics meets not only chemistry and biology but also
(nano-)technology.
\end{abstract}
\maketitle

\section{\label{sec1}Introduction}

Motility is the hallmark of life. What distinguishes a {\it traffic-like}
movement from all other forms of movements is that motile elements move 
on {\it ``tracks''} or {\it ``trails''}. However, in sharp contrast to 
vehicular traffic, the tracks and trails, which are the biological analogs 
of roads, can have nontrivial dependence on time during the typical travel 
time of the motile elements. What makes biological traffic even more 
unusual is that in many cases the motile elements themselves not only 
create the tracks but also modify their lengths as well as shape and, in 
some extreme cases, even leave behind a trail of destruction by wiping 
out the track as they move forward.

We are mainly interested in the {\it general principles} and common 
trends seen in the mathematical modeling of collective traffic-like 
movements at different levels of biological organization \cite{polrev}. 
Although the choice of the physical examples and modelling strategies 
are biased by our own works and experiences, we put these in a broader 
perspective by relating these with works of other research groups. We 
begin at the lowest level, starting with intracellular biomolecular 
motor traffic on filamentary rails. Then we present brief summaries of 
recent works on the traffic of molecular motors along the collagen 
fibrils in the extra-cellular matrix and those on transport of 
micron-size cargo by uni-cellular micro-organisms. We end our review by 
discussing the collective traffic-like terrestrial movements of social 
insects, particularly, ants, on their trails.

\section{Theoretical approaches}
\label{sec-approach}

In recent years many individual-based models of biological traffic 
have been formulated in discretized space. While in some models 
the dynamics of the system has been formulated in terms of differential 
equations assuming continuous time, in many others the dynamics 
has been implemented in terms of ``update rules'' in discrete time 
steps in the spirit of cellular automata and lattice gas models 
\cite{wolfram,chopard,marro}. Most of these models are extensions 
of a class of particle-hopping models which were earlier successfully 
used in the context of vehicular traffic \cite{css}.
The {\em asymmetric simple exclusion process (ASEP)} \cite{gunterrev}
is one of the simplest particle-hopping models. In the ASEP particles 
can hop (with some probability or rate) from one lattice site to a
neighbouring one, but only if the target site is not already occupied
by another particle.  ``Simple Exclusion'' thus refers to the absence
of multiply occupied sites. Generically, it is assumed that the motion
is ``asymmetric'' such that the particles have a preferred direction
of motion. 

For such driven diffusive systems the boundary conditions turn out
to be crucial. If periodic boundary conditions are imposed, i.e.,
the sites $1$ and $L$ are made nearest-neibours of each other, all
the sites are treated on the same footing. If the boundaries are open, 
then a particle can enter from a reservoir and occupy the leftmost 
site ($j=1$), with probability $\alpha$, if this site is empty. In 
this system a particle that occupies the rightmost site ($j=L$) can 
exit with probability $\beta$. The ASEP has been studied extensively 
in recent years and is now well understood \cite{gunterrev,evansAlten}. 
Both parallel and random-sequential updating schemes have been studied 
extensively in the literature.

The average number of motile elements that arrive at (or depart from) a
fixed detector site on the track per unit time interval is called the
{\it flux}. One of the most important transport properties is the relation
between the flux and the density of the motile elements; a graphical
representation of this relation is usually referred to as the {\it
fundamental diagram}. If the motile elements interact mutually only via
their steric repulsion their average speed $v$ would decrease with
increasing density because of the hindrance caused by each on the
following elements. On the other hand, for a given density $c$, the flux
$J$ is given by $J = c v(c)$, where $v(c)$ is the corresponding average
speed. At sufficiently low density, the motile elements are well separated
from each other and, consequently, $v(c)$ is practically independent of
$c$. Therefore, $J$ is approximately proportional to $c$ if $c$ is very
small. However, at higher densities the increase of $J$ with $c$ becomes
slower. At high densities, the sharp decrease of $v$ with $c$ leads to a
decrease, rather than increase, of $J$ with increasing $c$. Naturally, the
fundamental diagram of such a system is expected to exhibit a maxium at
an intermediate value of the density.

\section{Intra-cellular traffic of cytoskeletal molecular motors}

Intracellular transport is carried by molecular motors which are proteins
that can directly convert the chemical energy into mechanical energy
required for their movement along filaments constituting what is known
as the cytoskeleton \cite{howard,schliwa}. Three superfamilities of
these motors are kinesin, dynein and myosin; majority of these motors 
are two-headed.  Most of the kinesins and dyneins are like ``porters''
in the sense that these move over long distances carrying cargo along 
the filamentary tracks without getting completely detached; such motors 
are called {\it processive}. On the other hand, the conventional myosins 
and a few unconventional ones are nonprocessive; they are like ``rowers''. 

These cytoskeleton-based molecular motors play crucially important
biological functions, e.g., in axonal transport in neurons. The 
mechano-chemistry of single cytoskeletal motors and the mechanism 
of their motility have been investigated both experimentally and 
theoretically for quite some time \cite{osterrev,fisher,astu1}.


Often a single microtubule (MT) is used simultaneously by many motors 
and, in such circumstances, the inter-motor interactions cannot be 
ignored. In this article we shall focus mostly on the effects of mutual
interactions of these motors on their collective spatio-temporal 
organisation and the biomedical implications of such organisations. 
Fundamental understanding of these collective physical phenomena may 
also expose the causes of motor-related diseases (e.g., Alzheimer's 
disease) \cite{traffic,hirotaked,mandeld,goldstein} thereby helping, 
possibly, also in their control and cure. The bio-molecular motors have 
opened up a new frontier of applied research --- ``bio-nanotechnology''. 
A clear understanding of the mechanisms of these natural nano-machines 
will give us clue as to the possible design principles that can be 
utilized to synthesize artificial nanomachines.

Derenyi and collaborators \cite{derenyi1,derenyi2} developed
one-dimensional models of interacting Brownian motors. 
They modelled each motor as a {\it rigid rod}
and formulated the dynamics through Langevin equations for each such 
rod assuming the validity of the overdamped limit; the mutual 
interactions of the rods were incorporated through the mutual exclusion. 
        
The model considered by Aghababaie et al.\cite{menon1} is not based on
TASEP, but its dynamics is a combination of Brownian ratchet and update
rules in discrete time steps.  In this model the filamentary 
track is discretized and the motors are represented by {\it field-driven} 
particles in the spirit of the particle-hopping models. The hopping
probabilities of the particles are obtained from the instantaneous form 
of the local time-dependent potential. No site can accomodate more than one 
particle at a time. Each time step consists of either an attempt of a 
particle to hop to a neighbouring site or an attempt that can result in 
switching of the potential from flat to sawtooth form or vice-versa. 
Both forward and backward movement of the particles are possible. However, 
neither attachment of new particles nor complete detachment of existing 
particles were allowed.

The fundamental diagram of the model \cite{menon1}, computed imposing
periodic boundary conditions, is very similar to those of TASEP. This
observation indicates that further simplification of the model proposed
in ref.\cite{menon1} is possible to develope a minimal model for interacting
molecular motors. Indeed, the detailed Brownian ratchet mechanism, which
leads to a noisy forward-directed movement of the {\it field-driven}
particles in the model of Aghababaie et al. \cite{menon1}, is replaced
in some of the more recent theoretical models
\cite{lipo1,lipo7,lipo8,lipo9,frey1,frey2,santen1,santen2,popkov1}
by a TASEP-like probabilistic forward hopping of {\it self-driven}
particles \cite{schweitzer}. In these simplied versions, none of the 
particles is allowed
to hop backward and the forward hopping probability is assumed to capture
most of the effects of biochemical cycle of the enzymatic activity of the
motor. The explicit dynamics of the model is essentially an extension
of that of the asymmetric simple exclusion processes (ASEP)
\cite{sz,gunterrev} that includes, in addition, Langmuir-like kinetics 
of adsorption and desorption of the motors.

\begin{figure}[ht]
\begin{center}
\includegraphics[%
bb=9cm -1cm 19cm 30cm,
clip, angle=-90, width=10cm]{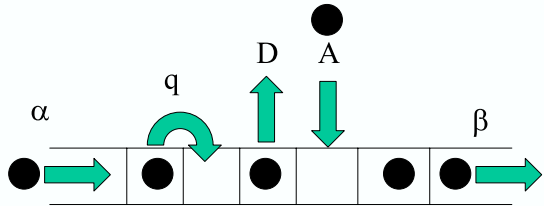}
\end{center}
\caption{A schematic description of the TASEP-like model introduced in
ref.~\cite{frey1} for molecular motor traffic. Just as in TASEP, the
motors are allowed to hop forward, with probability $q$. In addition,
the motors can also get ``attached'' to an empty lattice site, with
probability $A$, and ``detached'' from an occupied site, with probability
$D$ from any site except the end points; the rate of attachment at
the entry point on the left is $\alpha$ while that at the exit point
on the right is $\beta$.
}
\label{fig-frey}
\end{figure}

In the model of Parmeggiani et al. \cite{frey1}, the molecular motors
are represented by particles whereas the sites for the binding of the
motors with the cytoskeletal tracks (e.g., microtubules) are represented
by a one-dimensional discrete lattice.  Just as in TASEP, the motors are
allowed to hop forward, with probability $q$, provided the site in
front is empty. However, unlike TASEP, the particles can also get
``attached'' to an empty lattice site, with probability $A$,
and ``detached'' from an occupied site, with probability $D$
(see fig.\ref{fig-frey}) from any site except the end points. The state
of the system was updated in a random-sequential manner.
Carrying out Monte-Carlo simulations of the model, applying open
boundary conditions, Parmeggiani et al.\cite{frey1} demonstrated a
novel phase where low and high density regimes, separated from each
other by domain walls, coexist \cite{santen1,santen2}. Using a
mean-field theory (MFT), they interpreted this spatial organization
as traffic jam of molecular motors.

A cylindrical geometry of the model system was considered by Lipowsky, 
Klumpp and collaborators \cite{lipo1,lipo7,lipo8} to mimic the
microtubule tracks in typical tubular neurons. The microtubule filament
was assumed to form the axis of the cylinder whereas the free space
surrounding the axis was assumed to consist of $N_{ch}$ channels each
of which was discretized in the spirit of lattice gas models. They
studied concentration profiles and the current of free motors as well
as those bound to the filament by imposing a few different types of
boundary conditions. This model enables one to incorporate the effects
of exchange of populations between two groups, namely, motors bound to
the axial filament and motors which move diffusively in the cylinder.
They have also compared the results of these investigations with the
corresponding results obtained in a different geometry where the filaments
spread out radially from a central point.

\begin{figure}[ht]
\ \vspace{-1.3cm}
\begin{center}
\includegraphics[angle=-90,width=0.6\textwidth]{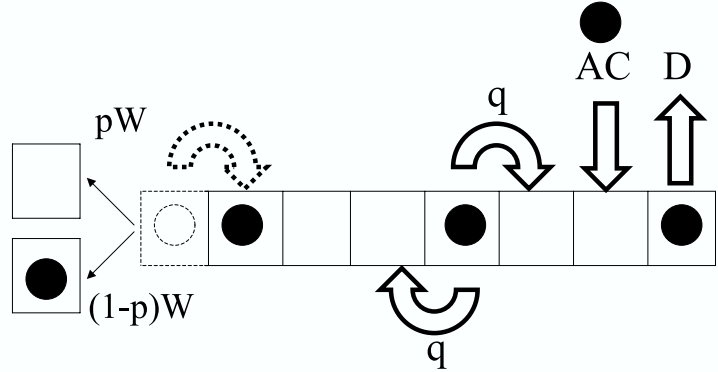}\\
\end{center}
\caption{A schematic description of the model suggested by Klein et al. 
\cite{klein} for motor induced depolymerization of cytoskeletal 
filaments. 
}
\label{fig-klein}
\end{figure}

A novel feature of the model of Klein et al. \cite{klein} 
(see Fig.~\ref{fig-klein}) is that the 
lattice site at the tip of a filament is removed
with a probability $W$ per unit time provided it is occupied by a motor;
the motor remains attached to the newly exposed tip of the filament
with probability $p$ (or remains bound with the removed site with
probability $1-p$). Thus, $p$ may be taken as a measure of the
processivity of the motors. This model clearly demonstrated a dynamic
accumulation of the motors at the tip of the filament arising from the
processivity. 

\subsection{Traffic of interacting single-headed motors KIF1A}
                                                                                 
The models of intracellular traffic described so far are essentially
extensions of the asymmetric simple exclusion processes (ASEP)
\cite{sz,gunterrev} that includes Langmuir-like kinetics of adsorption
and desorption of the motors. In reality, a motor protein is an enzyme
whose mechanical movement is loosely coupled with its biochemical cycle.
In a recent work \cite{nishietal}, we have considered specifically the
{\it single-headed} kinesin motor, KIF1A
\cite{okada1,okada2,okada3,Nitta,unpub};
the movement of a single KIF1A motor was modelled earlier with a Brownian
ratchet mechanism \cite{julicher,reimann}. In contrast to the earlier
models \cite{frey1,santen1,popkov1,lipo7} of molecular motor traffic,
which take into account only the mutual interactions of the motors, our
model explicitly incorporates also the Brownian ratchet mechanism of
individual KIF1A motors, including its biochemical cycle that involves
{\it adenosine triphosphate(ATP) hydrolysis}.
                                                                                 
\begin{figure}[ht]
\begin{center}
\includegraphics[width=0.75\textwidth]{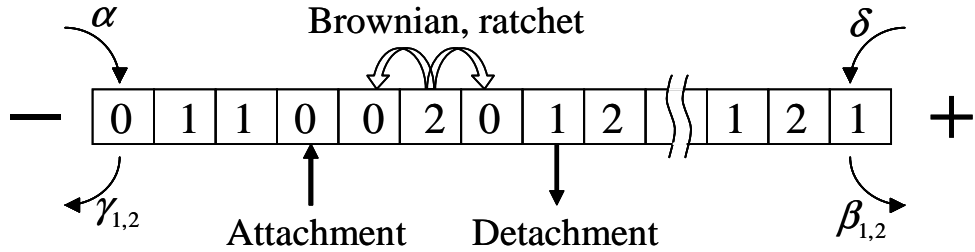}\\
\end{center}
\caption{A 3-state model for molecular motors moving along a MT.
0 denotes an empty site, 1 is K or KT and 2 is KD. Transition from
1 to 2, which corresponds to hydrolysis, occurs within a cell
whereas movement to the forward or backward cell occurs only when
motor is in state 2. At the minus and plus ends the probabilities
are different from those in the bulk.}
\label{fig2}
\end{figure}

The ASEP-like models successfully explain the occurrence of shocks.
But since most of the bio-chemistry is captured in these models through
a single effective hopping rate, it is difficult to make direct
quantitative comparison with experimental data which depend on such
chemical processes. In contrast, the model we proposed in ref.
\cite{nishietal} incorporates the essential steps in the biochemical
processes of KIF1A as well as their mutual interactions and involves
parameters that have one-to-one correspondence with experimentally
controllable quantities. Thus, in contrast to the earlier ASEP-like models, 
each of the self-driven
particles, which represent the individual motors KIF1A, can be in two
possible internal states labelled by the indices $1$ and $2$. In other
words, each of the lattice sites can be in one of three possible allowed
states (Fig.~\ref{fig2}): empty (denoted by $0$), occupied by a kinesin
in state $1$, or occupied by a kinesin in state $2$.

Good estimates for the parameters of the model could be extracted by
analyzing the empirical data \cite{nishietal}.
Assuming that each time step of updating corresponds to 1 ms of real
time, we performed simulations upto 1 minute. In the limit of low
density of the motors we have computed, for example, the mean speed of
the kinesins, the diffusion constant and mean duration of the movement
of a kinesin on a microtubule from simulations of our model; these are  
in excellent {\it quantitative} agreement with the corresponding 
empirical data from single molecule experiments.

Using this model we have also calculated the flux of the motors in the
mean field approximation imposing periodic boundary conditions.
Although the system with periodic boundary conditions is fictitious,
the results provide good estimates of the density and flux in the
corresponding system with open boundary conditions.
                                                                                 
In contrast to the phase diagrams in the $\alpha-\beta$-plane reported
by earlier investigators \cite{frey2,santen1,lipo7}, we have drawn the
phase diagram of our model in the $\omega_a-\omega_h$ plane by carrying 
out extensive computer simulations
for realistic parameter values of the model with open boundary conditions.
The phase diagram shows the strong influence of hydrolysis on the
spatial distribution of the motors along the MT. In particular, 
the position of the immobile shock depends on the concentration of the
motors as well as that of ATP; the shock moves towards the minus end of
the MT with the increase of the concentration of kinesin or ATP or both.
The formation of the shock has been established by our direct
experimental evidence; our findings on the domain wall are in qualitative 
agreement with the corresponding experimental observations \cite{nishietal}.

This work has been discussed in more detailed in our separate article 
\cite{nishicon} in this proceedings.

\section{Intra-cellular traffic of nucleotide-based motors} 

Helicases and polymerases are the two classes of nucleotide-based
motors that have been the main focus of experimental investigations.
In this section, we discuss only the motion of the ribosome along
the m-RNA track.  Historically, this problem is one of the first
where TASP-like model was successfully applied to a biological system.

\begin{figure}[h]
\begin{center}
\includegraphics[width=0.8\textwidth]{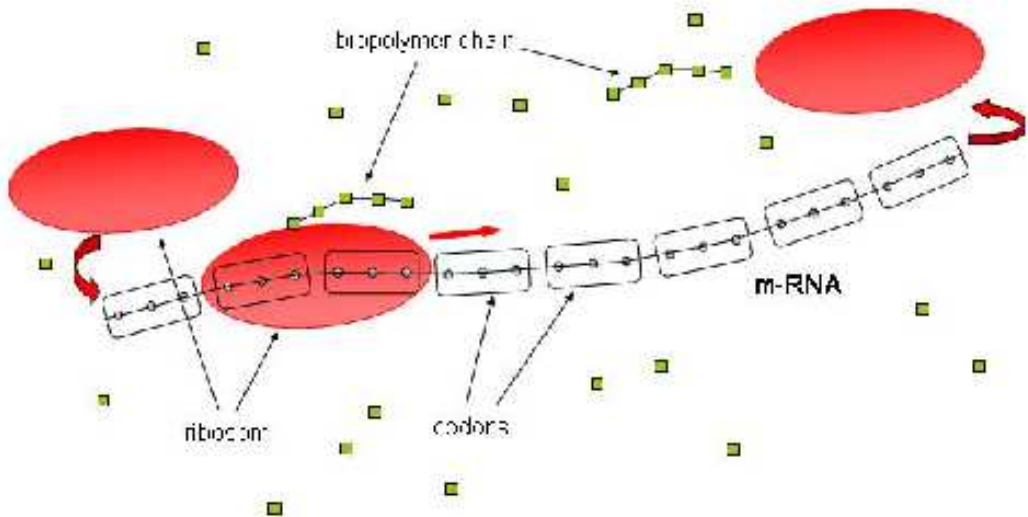}
\end{center}
\vspace{-0.6cm}
\caption{\protect{The process of biopolymerization: Ribosomes attach to
mRNA and read the construction plan for a biopolymer which is stored in
the genetic code formed by the sequence of codons. None of the codons
can be read by more than one ribosome simultaneously.
}}
\label{fig_mRNA}
\end{figure}

In a living cell {\em ribosomes} translate the genetic information 
`stored' in the {\em messenger-RNA (mRNA)} into a program for the 
synthesis of a protein. mRNA is a long (linear) molecule made up of a 
sequence of triplets of nucleotides; each triplet is called a {\it codon}
(see Fig.~\ref{fig_mRNA}). 
The genetic information is encoded in the sequence of codons. A ribosome, 
that first gets attached to the mRNA chain, ``reads'' the codons as it 
translocates along the mRNA chain, recruits the corresponding
amino acids and assembles these amino acids in the sequence thereby 
synthesizing the protein for which the ``construction plan'' was 
stored in the mRNA. Once the synthesis is completed, the ribosome 
gets detached from the mRNA. Thus, the process of ``translation'' of genetic
information stored in mRNA consists of three steps: (i) {\em initiation}:
attachment of a ribosome at the ``start'' end of the mRNA, (ii)
{\em elongation}: of the polypeptide (protein) as the ribosome moves along
the mRNA, and (iii) {\em termination}: ribosome gets detached from the
mRNA when it reaches the ``stop'' codon.

In order to model the traffic of ribosomes on a m-RNA track, let us 
denote each of the successive codons by the successive sites of
a one-dimensional lattice where the first and the last sites correspond
to the start and stop codons. The ribosomes are much bigger (20-30 times)
than the codons. Therefore, neighbouring ribosomes attached to the same
mRNA can not read the same information or overtake each other. In other
words, any given site on the lattice may be covered by a single ribosome
or none. Let us represent each ribosome by a rigid rod of length 
$\ell_{r}$. If the rod representing the  ribosome has its left edge  
attached to the i-th site of the lattice, it is allowed to move to the 
right by one lattice spacing, i.e., its left edge moves to the site $i+1$ 
provided the site $i+\ell_r$ is empty. In the special case $\ell_r = 1$ 
this model reduced to the TASEP. Although the model was originally proposed 
in the late sixties \cite{macdonald}, significant progress in its analytical
treatment for the general case of arbitrary $\ell_r$ could be made only 
three decades later; even the effects of quenched disorder has also been
considered in the recent literature
\cite{shaw1,shaw2,shaw3,lakatos1,lakatos2}.

As mentioned above, a ribosom is much bigger
than a base triplet. However, modifying the ASEP by taking into
account particles that occupy more than one lattice site does not
change the structure of phase diagram \cite{macdonald}.

\section{Extracellular transport: collagen-based motors}
\label{sec-extra}

The extracellular matrix (ECM) \cite{nagase} of vertebrates is rich in
collagen. Monomers of collagen form a triple-helical structure which
self-assemble into a tightly packed periodic organization of fibrils.
Cells residing in tissues can secret matrix metalloproteases (MMPs),
a special type of enzymes that are capable of degrading macromolecular
constituents of the ECM. The most notable among these enzymes is MMP-1
that is known to degrade collagen. The collagen fibril contains cleavage
sites which are spaced at regular intervals of $300$ nm. The collagenase
MMP-1 cleaves all the three $\alpha$ chains of the collagen monomer at
a single site.
                                                                                 
Breakdown of the ECM forms an essential step in several biological
processes like, for example, embryonic development, tissue remodelling,
etc. \cite{nagase}. Malfunctioning of MMP-1 has been associated with
wide range of diseases \cite{whittaker}. Therefore, an understanding
of the MMP-1 traffic on collagen fibrils can provide deeper insight
into the mechanism of its operation which, in turn, may give some clue
as to the strategies of control and cure of diseases caused by the
inappropriate functions of these enzymes.
                                                                                 
Saffarian et al. \cite{saffarian} used a technique of two-photon
excitation fluorescence correlation spectroscopy to measure the
correlation function corresponding to the MMP-1 moving along the
collagen fibrils. The measured correlation function strongly
indicated that the motion of the MMP-1 was not purely diffusive,
but a combination of diffusion and drift. In other words, the
``digestion'' of a collagen fibril occurs when a MMP-1 executes a
biased diffusion processively (i.e., without detachment) along
the fibril. They also demonstrated that inactivation of the
enzyme eliminates the bias but the diffusion remains practically
unaffected. They claimed that the energy required for the active
motor-like transport of the MMP-1 comes from the proteolysis (i.e.,
degradation) of the collagen fibrils.

\begin{figure}[t]
\begin{center}
\includegraphics[%
bb=2cm 2cm 11cm 30cm,
clip, angle=-90, width=10cm]{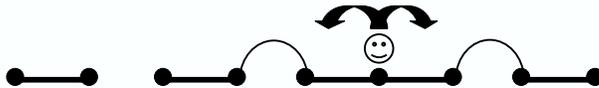}
\end{center}
\caption{Schematic representation of the one-dimensional burnt bridge 
model of MMP-1 dynamics proposed in ref.\cite{blumen}.
}
\label{fig-burnt1}
\end{figure}

There is a close relation between the traffic of MMP-1 on collagens 
and the ``burnt-bridge model'' introduced by Mai et al. \cite{blumen}.
In the burnt bridge model (see Fig.~\ref{fig-burnt1}), a 
``particle'' performs a random walk on a {\it semi-infinite}
{\it one-dimensional} lattice that extends from the origin to
$+\infty$. Each site of the lattice is connected to the two nearest
neighbour sites by links; a fraction $c$ of these links are called
``bridges'' and these are prone to be burnt by the random walker.
A bridge is burnt, with probability $p$, if the random walker either
crosses it {\it from left to right} or {\it attempts to cross if
from right to left} \cite{blumen,antal}. In either case, if the
bridge is actually burnt,
the walker stays on the right of the burnt bridge and cannot cross
it any more. The hindrance against leftward motion, that is created
by the burnt bridges, is responsible for the overall rightward drift
of the random walker.
Mai et al.\cite{blumen} studied the dependence of the average drift
velocity $v$ on the parameters $p$ and $c$ by computer simulation.
They also derived approximate analytical forms of these dependences
in the two limits $p \ll 1$ and $p \simeq 1$ using a continuum
approximation.

Saffarian et al.\ \cite{saffarian} also carried out computer
simulations of a two-dimensional model of the MMP-1 dynamics on
collagen fibrils which is essentially a two-dimensional generalization 
of the burnt-bridge model.  By comparing
the results of their simulations with their experimental observations,
Saffarian et al. they concluded that the observed biased diffusion
of the MMP-1 on collagen fibrils can be described quite well by a
Brownian ratchet mechanism \cite{julicher,reimann}.

\section{Cellular traffic}
\label{sec-cellular}
                                                                                 
A {\it Mycoplasma mobile} (MB) bacterium is an uni-cellular organism.
Each of the pear-shaped cells of this bacterium is about $700$ nm long
and has a diameter of about $250$ nm at the widest section. Each
bacterium can move fast on glass or plastic surfaces using a {\it
gliding} mechanism. In a recent experiment \cite{hiratsuka} narrow 
linear channels were constructed on lithographic substrates. The channels 
were typically $500$ nm wide and $800$ nm deep. Note that each channel 
was approximately twice as wide as the width of a single MB cell. 
The channels were so deep that none of the individual MB cells was able 
to climb up the tall walls of the channels and continued moving along 
the bottom edge of the walls of the cannels. In the absence of direct 
contact interaction with other bacteria, each individual MB cell was 
observed to glide, without changing direction, at an average speed of 
a few microns per second.
                                                                 
When two MB cells made a contact approaching each other from opposite
directions within the same channel, one of the two cells gave way and
moved to the adjacent lane. However, in a majority of the cases, two
cells approaching each other from the opposite directions simply passed
by as if nothing had happened; this is because of the fact that the
width of the channel is roughly twice that of the individual MB cell.
Moreover, when two cells moving in the same direction within a channel
collided with each other, the faster cell moved to the adjacent lane
after the collision.
                                                                                 
Hiratsuka et al.\cite{hiratsuka} attached micron-sized beads on the
MB cells using biochemical technique and demonstrated that the average
speed of each MB cell remained practically unaffected by the load it
was carrying. In contrast to the nonliving motile elements discussed
in all the preceedings sections, the cells are the functional units of
life. Therefore, the MB cells have the potential for use in applied
research and technology as ``micro-transporters''. More recently, the
unicellular biflagellated algae {\it Chlamydomonas reinhardtii} (CR),
which are known to be phototactic {\it swimmers}, have been shown to
be even better candidates as ``micro-transporter'' as these can attain
average speeds that is about two orders of magnitude higher than what
was possible with MB cells \cite{weibel}. However, to our knowledge,
the effects of mutual interactions of the CR cells on their average
speed at higher densities has not been investigated.

\section{Traffic in social insect colonies: ants and termites}
\label{sec-ants}
                                                                                 
From now onwards, we shall study traffic of
multi-cellular organisms, particularly, ants which are social insects. 
 The
ability of the social insect colonies to function without a leader
has attracted the attention of experts from various disciplines
\cite{bonabu97,anderson02,huang,bonabu98,theraulaz03,gautrais,keshet94,theraulazetal}. 
Insights gained from the modeling of the
colonies of such insects are finding important applications
in computer science (useful optimization and control algorithms)
\cite{dorigo}, communication engineering \cite{bona00}, artificial
``swarm intelligence'' \cite{bonabeau} and micro-robotics \cite{krieger}
as well as in task partitioning, decentralized manufacturing
\cite{anderson99a,anderson99b,anderson99c,anderson00a,anderson01,anderson00b}
and management \cite{meyer}.
the collective terrestrial movements
of ants have close similarities with the other traffic-like phenomena
considered here. When observed from a sufficiently long distance the
movement of ants on trails resemble the vehicular traffic observed from
a low flying aircraft.

Ants communicate with each other by dropping a chemical (generically
called {\it pheromone}) on the substrate as they move forward
\cite{wilson,camazine,mikhailov}. Although we cannot smell it, the
trail pheromone sticks to the substrate long enough for the other
following sniffing ants to pick up its smell and follow the trail.
\cite{wilson}. Both the continuum model developed by Rauch et al.\cite{rauch} 
and the
CA model introduced by Watmough and Edelstein-Keshet \cite{watmough}
were intended to address the question of formation of the ant-trail
networks by foraging ants.
Couzin and Franks \cite{couzin} developed an individual based model
that not only addressed the question of self-organized lane formation
but also elucidated the variation of the flux of the ants. 

In the recent years, we have developed discrete models 
\cite{cgns,ncs,kunwar,jscn} that are not intended to address the 
question of the emergence of the ant-trail \cite{activewalker}, but 
focus on the traffic of ants on a trail which has already been formed. 
We have developed models of both unidirectional and bidirectional 
ant-traffic by generalizing the totally asymmetric simple exclusion 
process (TASEP) \cite{derrida1,derrida2,gunterrev} with parallel 
dynamics by taking into account the effect of the pheromone.

\begin{figure}[ht]
\begin{center}
\includegraphics[width=0.65\textwidth]{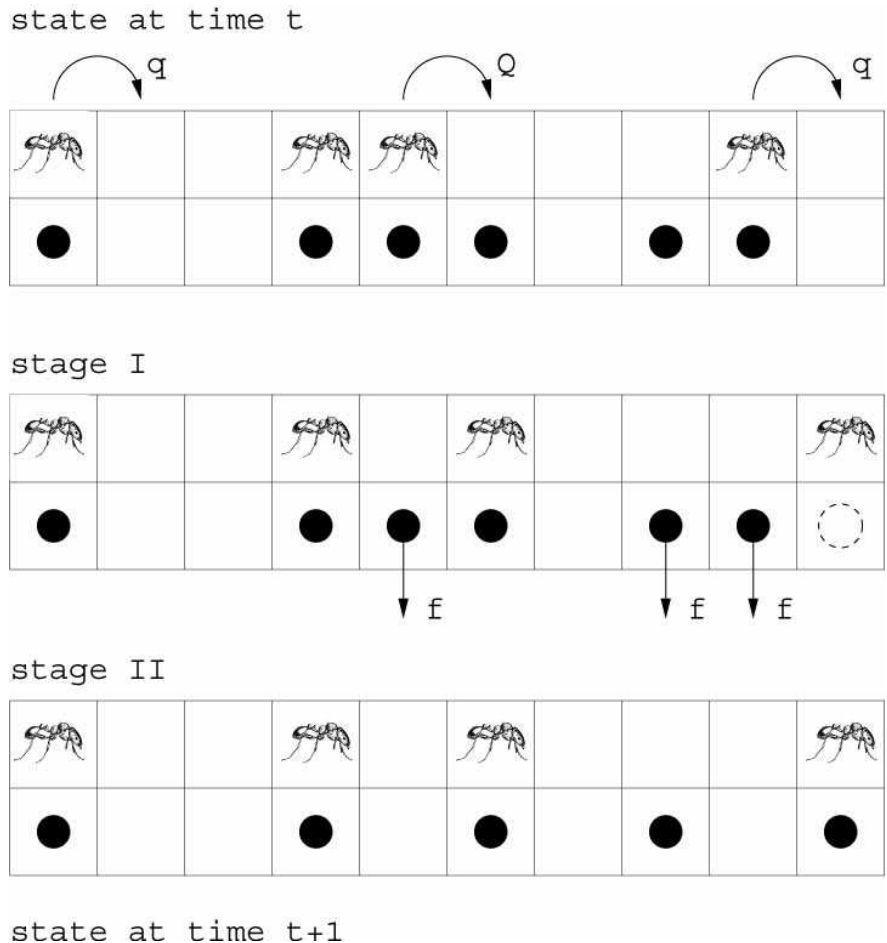}
\end{center}
\caption{
Schematic representation of typical configurations of the
uni-directional ant-traffic model. The symbols $\bullet$ indicate
the presence of pheromone.
This figure also illustrates the update procedure.
Top: Configuration at time $t$, i.e.\ {\it before} {\em stage I}
of the update. The non-vanishing probabilities of forward movement of
the ants are also shown explicitly. Middle: Configuration {\it after}
one possible realisation of {\it stage I}. Two ants have moved compared
to the top part of the figure. The open circle with dashed boundary
indicates the location where pheromone will be dropped by the corresponding
ant at {\em stage II} of the update scheme. Also indicated are the existing
pheromones that may evaporate in {\em stage II} of the updating, together
with the average rate of evaporation.  Bottom: Configuration {\it after}
one possible realization of {\it stage II}. Two drops of pheromones
have evaporated and pheromones have been dropped/reinforced at the
current locations of the ants.
}
\label{fig-modeldef1}
\end{figure}

In our model of uni-directional ant-traffic the ants move according to
a rule which is essentially an extension of the TASEP dynamics.
In addition, a second field is introduced which models
the presence or absence of pheromones (see Fig.~\ref{fig-modeldef1}).
The hopping probability of the ants is now modified by the presence of
pheromones. It is larger if a pheromone is present at the destination site.
Furthermore, the dynamics of the pheromones has to be specified. They
are created by ants and free pheromones evaporate with probability $f$
per unit time. Assuming periodic boundary conditions, the state of the
system is updated at each time step in two stages. 
In stage I ants are allowed to move
while in stage II the pheromones are allowed to evaporate. In each
stage the {\it stochastic} dynamical rules are applied in parallel to
all ants and pheromones, respectively.\\

One interesting phenomenon observed in the simulations is coarsening.
At intermediate time usually several non-compact clusters are formed.
However, the velocity of a cluster depends on the distance to the next 
cluster ahead. Obviously, the probability that the pheromone created 
by the last ant of the previous cluster survives decreases with 
increasing distance. Therefore clusters with a small headway move 
faster than those with a large headway. This induces a coarsening 
process such that after long times only one non-compact cluster 
survives.

A similar coarsening phenomenon has been observed also in the bus-route 
model \cite{loan,cd}. In fact, the close relation between our model of 
uni-directional traffic on ant-trails and the bus-route model has been 
pointed out earlier \cite{kunwar2}. In the bus route model, each bus 
stop can accomodate at most one bus at a time; the passengers arrive at 
the bus stops randomly at an average rate $\lambda$ and each bus, which 
normally moves from one stop to the next at an average rate $Q$, slows 
down to $q$, to pick up waiting passengers  \cite{loan,cd}. 

In vehicular traffic, usually, the average speed of the vehicles decreases 
{\it monotonically} with increasing density because the inter-vehicle 
interactions tend to hinder each other's movements. In contrast, in our 
models of ant-traffic the average speed of the ants varies {\it 
non-monotonically} with their density over a wide range of small values 
of $f$ because of the coupling of their dynamics with that of the 
pheromone. This uncommon variation of the average speed gives rise to 
the unusual dependence of the flux on the density of the ants in our 
models of ant-traffic.  Furthermore, the flux does not exhibit the 
particle-hole symmetry which is a characteristic of the TAEP. 
Details of the models and results on both uni-directional and bi-directional 
traffic of ants on trails are given in the article of John et al. in this 
proceedings \cite{johnetal}. The experimental data reported in the 
pioneering experimental work of Burd and collaborators \cite{burd1} 
were too scattered to test our theoretical predictions. However, more 
accurate recent data \cite{burd2,john2} establish both the non-monotonic 
variation of the average speed with density as well as the formation of 
cluster by the ants.

\section{Summary and conclusion}
\label{sec-sum}

In this article we have reviewed our current understanding of traffic-like
collective phenomena in living systems, starting from the smallest level
of intra-cellular bio-molecular motor transport and ending at the level 
of the traffic of social insects like, for example, ants. We have 
restricted our attention to those theoretical works where, in the spirit 
of particle-hopping models of vehicular traffic, the language of cellular 
automata or extensions of TASEP has been used. The success of this modelling
strategy has opened up a new horizon in traffic science and, we hope, we 
have provided a glimpse of an exciting frontier of interdisciplinary 
research.

\vspace{1cm}
                                                                                
\noindent {\bf Acknowledgements:} It is our great pleasure to thank
Yasushi Okada, Alexander John and Ambarish Kunwar for enjoyable
collaborations on the topics discussed in this article. 


\begin{thebibliography}{99}
          
\bibitem{polrev} D. Chowdhury, A. Schadschneider and K. Nishinari, 
Phys. of Life Rev. (2005) in press.                                
\bibitem{wolfram} S. Wolfram, {\it Theory and Applications of Cellular
Automata} (World Sci., 1986); {\em A New Kind of Science}
(Wolfram Research Inc., 2002)
\bibitem{chopard} B. Chopard and M. Droz, {\it Cellular Automata Modeling
of Physical Systems} (Cambridge University Press, 1998).
\bibitem{marro} J. Marro and R. Dickman, {\it Nonequilibrium Phase
Transitions in Lattice Models} (Cambridge University Press, 1999).
\bibitem{css} D. Chowdhury, L. Santen, and A. Schadschneider, Phys. Rep.
{\bf 329}, 199 (2000). 
\bibitem{gunterrev} G.M. Sch\"utz: {\em Exactly Solvable Models for
Many-Body Systems}, in C.\ Domb and J.L.\ Lebowitz (eds.),
{\em Phase Transitions and Critical Phenomena}, Vol.~19
(Academic Press, 2001).
\bibitem{evansAlten} M.R.\ Evans and R.A.\ Blythe,
Physica {\bf A313}, 110 (2002).
\bibitem{howard} J. Howard, {\it Mechanics of Motor Proteins and the
Cytoskeleton}, (Sinauer Associates, 2001) .
\bibitem{schliwa} M. Schliwa (ed.), {\it Molecular Motors}, (Wiley-VCH, 2002).
\bibitem{osterrev} G. Oster and H. Wang, in ref.\cite{schliwa}.
\bibitem{fisher} M.E. Fisher and A.B. Kolomeisky, Proc. Natl. Acad. Sci.
{\bf 98}, 7748 (2001).
\bibitem{astu1} R.D. Astumian, Appl. Phys. A {\bf 75}, 193 (2002).
\bibitem{traffic} M. Aridor and L.A. Hannan, Traffic {\bf 1}, 836 (2000);
{\bf 3}, 781 (2002).
\bibitem{hirotaked} N. Hirokawa and R. Takemura,
Trends in Biochem. Sci. {\bf 28}, 558 (2003)
\bibitem{mandeld} E. Mandelkow and E.M. Mandelkow, Trends in Cell Biol.
{\bf 12}, 585 (2002).
\bibitem{goldstein} L.S. Goldstein, Proc. Natl. Acad. Sci. {\bf 98}, 6999
(2001); Neuron {\bf 40}, 415-425 (2003).
{\bf 28}, 558 (2003);  Curr. Op. Neurobiol. {\bf 14}, 564-573 (2004).
\bibitem{derenyi1} I. Derenyi and T. Vicsek, Phys. Rev. Lett. {\bf 75}, 374
(1995).
\bibitem{derenyi2} I. Derenyi and A. Ajdari, Phys. Rev. E {\bf 54}, R5 (1996).
\bibitem{menon1} Y. Aghababaie, G.I. Menon and M. Plischke, Phys. Rev. E
{\bf 59}, 2578 (1999).
\bibitem{lipo1} R. Lipowksy, S. Klumpp, and Th. M. Nieuwenhuizen, Phys. Rev.
Lett. 87, 108101 (2001).
\bibitem{lipo7} R. Lipowksy and S. Klumpp, Physica A 352, 53 (2005).
\bibitem{lipo8} M.J.I. M\"uller, S, Klumpp and R. Lipowsky, J. Phys. Cond. Matt. {\bf 17}, S3839 (2005) and references therein. 
\bibitem{lipo9} S. Klumpp and R. Lipowsky, this proceedings.
\bibitem{frey1}  A. Parmeggiani, T. Franosch, and E. Frey, Phys. Rev. Lett.
{\bf 90}, 086601 (2003); Phys. Rev. E {\bf 70}, 046101 (2004).
\bibitem{frey2} E. Frey, A. Parmeggiani and T. Franosch, Genome Informatics {\bf 15(1)}, 46 (2004) and references therein.
\bibitem{santen1} M.R. Evans, R. Juhasz, and L. Santen, Phys. Rev. E {\bf 68},
026117 (2003).
\bibitem{santen2} R. Juhasz and L. Santen, J.~Phys.\ A {\bf 37}, 3933 (2004).
\bibitem{popkov1} V. Popkov, A. Rakos, R.D. Williams, A.B. Kolomeisky, and
G.M. Sch\"utz, Phys. Rev. E {\bf 67}, 066117 (2003).
\bibitem{schweitzer} F.\ Schweitzer: {\em Brownian Agents and Active
Particles}, Springer Series in Synergetics (Springer 2003).
\bibitem{sz} B. Schmittmann and R.P.K.\ Zia,
in C.\ Domb and J.L.\ Lebowitz (eds.),
{\em Phase Transitions and Critical Phenomena}, Vol.~17
(Academic Press, 1995).
\bibitem{klein} G.A. Klein, K. Kruse, G. Cuniberti and F. J\"ulicher,
Phys Rev. Lett. {\bf 94}, 108102 (2005).
\bibitem{nishietal} K. Nishinari, Y. Okada, A. Schadschneider and
D. Chowdhury, Phys. Rev. Lett. {\bf 95}, 118101 (2005).
\bibitem{okada1} Y. Okada and N. Hirokawa, Science {\bf 283}, 1152 (1999).
\bibitem{okada2} Y. Okada and N. Hirokawa, Proc. Natl. Acad.Sci. USA
{\bf 97}, 640 (2000).
\bibitem{okada3} Y. Okada, H. Higuchi, and N. Hirokawa, Nature, {\bf 424},
574 (2003).
\bibitem{Nitta} R. Nitta, M. Kikkawa, Y. Okada, and N. Hirokawa,
  Science {\bf 305}, 678 (2003).
\bibitem{unpub} Y. Okada, K. Nishinari, D. Chowdhury, A. Schadschneider,
and N. Hirokawa (to be published).
\bibitem{julicher} F. J\"ulicher, A. Ajdari, and J. Prost,
Rev. Mod. Phys. {\bf 69}, 1269 (1997).
\bibitem{reimann} P. Reimann, Phys. Rep. {\bf 361}, 57-265 (2002).
\bibitem{nishicon} K. Nishinari, Y. Okada, A. Schadschneider and D. Chowdhury, 
in this proceedings.
\bibitem{macdonald} C. MacDonald, J. Gibbs, and A. Pipkin, Biopolymers
{\bf 6}, 1 (1968); C. MacDonald and J. Gibbs, Biopolymers {\bf 7}, 707 (1969)
\bibitem{shaw1} L.B. Shaw, R.K.P. Zia and K.H. Lee, Phys. Rev. E {\bf 68},
021910 (2003).
\bibitem{shaw2} L.B. Shaw, J. P. Sethna and K.H. Lee, Phys. Rev. E {\bf 70},
021901 (2004).
\bibitem{shaw3} L.B. Shaw, A.B. Kolomeisky and K.H. Lee, J. Phys. A {\bf 37},
2105 (2004).
\bibitem{lakatos1} G. Lakatos and T. Chou, J. Phys. A {\bf 36}, 2027 (2003).
\bibitem{lakatos2} T. Chou and G. Lakatos, Phys. Rev. Lett. {\bf 93}, 198101
(2004).
\bibitem{nagase} H. Nagase and J. F. Woessner, J. Biol. Chem. {\bf 274},
21491 (1999).
\bibitem{whittaker} M. Whittaker and A. Ayscough,
Celltransmisions {\bf 17}, 3 (2001).
\bibitem{saffarian} S. Saffarian, I. E. Collier, B.L. Marmer, E.L. Elson and
G. Goldberg, Science {\bf 306}, 108 (2004).
\bibitem{blumen} J. Mai, I.M. Sokolov and A. Blumen, Phys. Rev. E {\bf 64},
011102 (2001).
\bibitem{antal} T. Antal and P.L. Krapivsky, cond-mat/0504652.
\bibitem{hiratsuka} Y. Hiratsuka, M. Miyata and T. Q. P. Uyeda,
Biochem. Biophys. Res. Commun. {\bf 331}, 318 (2005).
\bibitem{weibel} D. B. Weibel, P. Garstecki, D. Ryan, W. R. DiLuzio, M. Mayer, J. E. Seto and G. M. Whitesides, Proc. Nat. Acad. Sci. USA, {\bf 102}, 11963 (2005).
\bibitem{bonabu97} E. Bonabeau, G. Theraulaz, J.L. Deneubourg, S. Aron and
S. Camazine,
Trends in Ecol. Evol. {\bf 12}, 188 (1997)
\bibitem{anderson02} C. Anderson, G. Theraulaz and J.L. Deneubourg,
Insect. Sociaux {\bf 49}, 99 (2002)
\bibitem{huang} Z. Huang and J.H. Fewell,
Trends in Ecol. Evol.  {\bf 17}, 403 (2002).
\bibitem{bonabu98} E. Bonabeau, Ecosystems {\bf 1}, 437 (1998).
\bibitem{theraulaz03} G. Theraulaz, J. Gautrais, S. Camazine and
J.L. Deneubourg,
Phil. Trans. Roy. Soc. Lond. A {\bf 361}, 1263 (2003).
\bibitem{gautrais} J. Gautrais, G. Theraulaz, J.L. Deneubourg and
C. Anderson,
J. Theor. Biol. {\bf 215}, 363 (2002).
\bibitem{keshet94} L. Edelstein-Keshet,
J. Math. Biol. {\bf 32}, 303 (1994).
\bibitem{theraulazetal} G. Theraulaz, E. Bonabeau, S.C. Nicolis, R.V. Sole,
V. Fourcassie, S. Blanco, R. Fournier, J.L. Joly, P. Fernandez, A. Grimal,
P. Dalle and J.L. Deneubourg,
Proc. Natl.Acad. Sci. {\bf 99}, 9645 (2002).
\bibitem{dorigo} M. Dorigo, G. di Caro and L.M. Gambardella,
Artificial Life {\bf 5(3)}, 137 (1999); Special issue of Future
Generation Computer Systems dedicated to ant-algorithms (2000).
\bibitem{bona00} E. Bonabeau, M. Dorigo and G. Theraulaz,
Nature {\bf 400}, 39 (2000).
\bibitem{bonabeau} E. Bonabeau, M. Dorigo and G. Theraulaz,
{\it Swarm Intelligence: From Natural to Artificial Intelligence}
(Oxford University Press, 1999).
\bibitem{krieger} M.J.B. Krieger, J.B. Billeter and L. Keller,
Nature {\bf 406}, 992 (2000).
\bibitem{anderson99a} F.L.W. Ratnieks and C. Anderson,
Insectes Sociaux {\bf 46}, 95 (1999).
\bibitem{anderson99b} C. Anderson and F.L.W. Ratnieks,
Am. Nat. {\bf 154}, 521 (1999).
\bibitem{anderson99c} F.L.W. Ratnieks and C. Anderson,
Am. Nat. {\bf 154}, 536 (1999).
\bibitem{anderson00a} C. Anderson and F.L.W. Ratnieks,
Insectes Sociaux {\bf 47}, 198 (2000).
\bibitem{anderson01}  C. Anderson and D.W. McShea,
Biol. Rev. {\bf 76}, 211 (2001).
\bibitem{anderson00b}  C. Anderson and F.L.W. Ratnieks,
in: {\it Complexity and complex systems in industry}, eds. I.P. McCarthy
and T. Rakotobe-Joel, (University of Warwick, U.K.), 92 (2000).
\bibitem{meyer} E. Bonabeau and C. Meyer,
Harvard Business Review (May), 107 (2001).
\bibitem{wilson} E.O. Wilson, {\it The Insect Societies}
(Belknap, Cambridge, USA, 1971);
B. H\"olldobler and E.O. Wilson, {\it The Ants} (Belknap, Cambridge, USA, 1990)
\bibitem{camazine} S. Camazine, J.L. Deneubourg, N. R. Franks,
J. Sneyd, G. Theraulaz, E. Bonabeau:
{\it Self-organization in Biological Systems} (Princeton
University Press, 2001).
\bibitem{mikhailov} A.S. Mikhailov and V. Calenbuhr, {\it From Cells
    to Societies: Models of Complex Coherent Action} (Springer, 2002).
\bibitem{rauch} E.M. Rauch, M. M. Millonas and D.R. Chialvo,
Phys. Lett. A {\bf 207}, 185 (1995).
\bibitem{watmough}  J. Watmough and L. Edelstein-Keshet,
J. Theor. Biol. {\bf 176}, 357 (1995).
\bibitem{couzin} I.D. Couzin and N.R. Franks, Proc. Roy Soc. London B
{\bf 270}, 139 (2003).
\bibitem{cgns} D. Chowdhury, V. Guttal, K. Nishinari, A. Schadschneider,
J. Phys. A:Math. Gen. {\bf 35}, L573 (2002)
\bibitem{ncs} K.\ Nishinari, D.\ Chowdhury, A.\ Schadschneider,
Phys.\ Rev.\ E {\bf 67}, 036120 (2003)
\bibitem{kunwar} A. Kunwar, D. Chowdhury, A. Schadschneider and K. Nishinari,
submitted for publication.
\bibitem{jscn}  A. John, A. Schadschneider, D. Chowdhury and K. Nishinari,
J. Theor. Biol. {\bf 231}, 279 (2004).
\bibitem{activewalker} D.\ Helbing, F.\ Schweitzer, J.\ Keltsch, P.\ Molnar:
Phys.\ Rev.\ {\bf E56}, 2527 (1997)
\bibitem{derrida1} B. Derrida, Phys. Rep. {\bf 301}, 65 (1998)
\bibitem{derrida2} B. Derrida and M.R. Evans, in: {\it Nonequilibrium
Statistical Mechanics in One Dimension}, ed. V. Privman
(Cambridge University Press, 1997)
\bibitem{loan} O.J.\ O'Loan, M.R.\ Evans, M.E.\ Cates,
Europhys. Lett. {\bf 42}, 137 (1998);
Phys. Rev. E{\bf 58}, 1404 (1998).
\bibitem{cd}  D.\ Chowdhury, R.C.\ Desai,
Eur. Phys. J. B{\bf 15}, 375 (2000). 
\bibitem{kunwar2} A. Kunwar, A. John, K. Nishinari, A. Schadschneider and 
D. Chowdhury, J. Phys. Soc. Jap. {\bf 73}, 2979 (2004).
\bibitem{johnetal} A. John, A. Kunwar, A. Namazi, A. Schadschneider,
D. Chowdhury, and K. Nishinari, in this proceedings.
\bibitem{burd1} M. Burd, D. Archer, N. Aranwela and D.J. Stradling,
Am. Nat. {\bf 159}, 283 (2002).
\bibitem{burd2} M. Burd et al. (2005) unpublished.
\bibitem{john2} A. John et al. (2005) unpublished.


\end{thebibliography}
\end{document}